\documentclass[10pt,onecolumn]{article}
\usepackage{graphicx}
\usepackage{amssymb}
\usepackage{amsmath}
\usepackage{upgreek}
\usepackage{breqn}
\usepackage{multirow}

\usepackage[labelfont=bf,labelsep=period,footnote size]{caption}
\usepackage{color}
\usepackage[
 breaklinks=true
 ,colorlinks=true
 ,linkcolor=blue 
 ,citecolor=blue
 ,backref=none
 ,pagebackref=false
 ]{hyperref} 

\usepackage{setspace}
\usepackage{epstopdf}
\pdfoutput=1
\usepackage{geometry}
\usepackage[auth-sc-lg,affil-sl]{authblk}

\usepackage{setspace} 
\usepackage[mathlines]{lineno}
\usepackage{todonotes}

\makeatletter
\renewcommand{\@biblabel}[1]{\quad#1.}
\makeatother

\newcommand{\be}{\begin{equation}}
\newcommand{\ee}{\end{equation}}
\newcommand{\bd}{\begin{displaymath}}
\newcommand{\ed}{\end{displaymath}}
\newcommand{\BE}{\begin{eqnarray}}
\newcommand{\EE}{\end{eqnarray}}

\newcommand{\id}{{\rm 1\!\!I}}

\newcommand{\mumat}{\underline{\underline{\mu}}}

\newcommand{\A}[1]{A}
\newcommand{\B}[1]{B}

\graphicspath{ {images/} }

\begin{document}


\title{Sojourn times and fixation dynamics in multi-player games with fluctuating environments}

\author{Joseph W. Baron\thanks{joseph.baron@student.manchester.ac.uk}}
\author{Tobias Galla \thanks{tobias.galla@manchester.ac.uk}}
\affil{Theoretical Physics, School of Physics and Astronomy, \\The University of Manchester, Manchester M13 9PL, United Kingdom}

\label{firstpage} 


\maketitle


\begin{abstract}
We study evolutionary multi-player games in finite populations, subject to fluctuating environments. The population undergoes a birth-death process with absorbing states, and the environment follows a Markovian process, resulting in a fluctuating payoff matrix for the evolutionary game. Our focus is on the fixation or extinction of a single mutant in a population of wildtypes. We show that the nonlinear nature of fitnesses in multi-player games gives rise to an intricate interplay of selection, genetic drift and environmental fluctuations. This generates effects not seen in simpler two-player games. To analyse trajectories towards fixation we analytically calculate sojourn times for general birth-death processes in populations of two types of individuals and in fluctuating environments. 
\end{abstract}

\onehalfspacing
\section{Introduction}
Models of evolutionary dynamics frequently involve randomness, and the timing of birth, death and mutation events is statistical. The modelling framework most commonly used in evolutionary dynamics, game theory, epidemiology and  population dynamics is that of a Markovian birth-death process, see e.g. \cite{traulsen:bookchapter:2009,nowak:book:2006}. These processes capture so-called intrinsic stochasticity (or demographic noise) in finite populations. More traditional approaches (see e.g. \cite{taylor:MB:1978,hofbauer:book:1998}), based on deterministic rate equations, neglect all randomness and are valid formally only in the limit of infinite populations. Deterministic models are analysed relatively easily using tools from nonlinear dynamics, however they do not capture fluctuation-driven phenomena such as fixation and extinction. These features can only be characterised within a stochastic model, and the analysis is frequently based on methods from non-equilibrium statistical mechanics \cite{gardinerbook,kampenbook}.

Often the relative growth rates or interaction coefficients between the different types of individuals themselves fluctuate in time, as environmental conditions change. Fluctuating environments lead to a second layer of stochasticity, in addition to the intrinsic stochasticity in finite populations. Examples of fluctuating environments include variation of external conditions (e.g., temperature, pH, presence or absence of nutrients), or targeted intervention, such as phases of antibiotic treatment, see e.g. \cite{thattai:2004, acar:NG:2008, kussell_2005, leibler:PNAS:2010, kussell:Science:2005}. Biological examples include the dynamics of persister cells in bacterial populations \cite{kussell_2005,patra:PLOS}, and the emergence of phenotype switching \cite{patra:BP}. Theoretical work exists to study such systems \cite{thattai:2004,patra:BP, ganter:2007,melbinger}, but often disregards intrinsic stochasticity or focuses on growth in continuously varying environments. We are mostly interested in the dynamics of fixation in fluctuating environments, which is not accounted for in the deterministic approach.

Previous work \cite{ashcroft} addressed the case of two-player evolutionary dynamics in switching environments with discrete states, and an analytical framework was constructed to calculate fixation probabilities and times. The analysis focused on games with linear payoffs and two-player interaction. Even for this relatively simple case, unexpected behaviour was observed due to the environmental switching; for example fixation probabilities of an invading mutant may be maximised for intermediate switching rates of the environmental process. 

The purpose of this work is to study fixation of mutants in evolutionary multi-player games in fluctuating environments. In multi-player games each individual interacts with more than one other individual in each instance of the game, and the resulting payoff or fitness functions become non-linear in the composition of the population. This can generate multiple non-trivial fixed points of the resulting replicator equations \cite{gamesmultiverse}; these are the equilibria of selection at which different species coexist and have the same fitness. These equilibrium points shape the dynamics and outcome of evolution in the presence of demographic noise. We are interested in the interplay between selection, random drift and environmental fluctuations in this non-linear scenario. We approach this in a general setting and in the context of a stylised minimal model. While we do not aim to model an immediate application, we believe our approach is the most suitable to identify the main phenomena and principal mechanisms. On the other hand it is clear that the scenario of multi-player interaction is of importance for a number of applications, and potentially more prevalent in the real world than two-player games. The latter are often analysed as an approximation as they are mathematically easier to handle. Applications of multi-player games have extensively been surveyed for example in \cite{gamesreview, broom2}. These include modelling in ecology (e.g. biological markets and auctions \cite{markets1, markets2, auctions}), population genetics (e.g dynamics of the Medea allele, and `playing the field' in the context of the sex-ratio game \cite{broom, hamilton}), and in the social sciences (e.g. public good games \cite{tragedy, scheuring}).

Multi-player games in finite populations have of course been investigated in the literature (see e.g. \cite{du:JRSI:2014,wu}). However, most existing studies focus on one fixed multi-player game, whereas our aim is to analyse  cases in which the game is shaped by an external fluctuating environment. In order to do this, we develop a theory for the analytical calculation of sojourn times for arbitrary two-species birth-death processes in randomly switching environments. We then apply it to understand fixation dynamics in fluctuating three-player games, and study the success of an invading mutant in an existing population of wildtypes.

The remainder of this paper is organised as follows. In Sec. \ref{sec:model} we define the components of the model, in particular the birth-death dynamics, the environmental process and the setup of a multi-player game. We introduce sojourn times for general birth-death processes in switching environments in Sec. \ref{sec:sojourn}; these can be calculated analytically, full mathematical details can be found in the Supplement. We then study fixation phenomena in three-player games in Sec. \ref{sec:3playergames}, before we present out conclusions and an outlook in Sec. \ref{sec:outlook}. Further technical details of the model setup and the analysis can be found in the Supplement. 

\section{Model definition}\label{sec:model}
\subsection{Evolutionary dynamics and environmental switching}
We consider a population of $N$ individuals. The size of the population does not vary in time. Each individual can be of one of two types, $A$ or $B$; we will occasionally refer to these as `species'. The population is assumed to be well mixed, i.e., every individual in the population can interact with any other individual. The state of the population is therefore fully specified by the number of individuals of type $A$, which we write as $i\in\{0,\dots,N\}$. The number of individuals of type $B$ is $N-i$. Evolution occurs through a discrete-time birth-death process, governed by transition probabilities $\omega_i^\pm$ from state $i$ to states $i\pm 1$, respectively. More precisely, $\omega_i^+$ is the probability to find the population in state $i+1$ in the next time step, if it is currently in state $i$. A similar definition applies for $\omega_i^-$, and $1-\omega_i^+-\omega_i^-$ is the probability for the population to remain in $i$. 

To model environmental influence, we assume that the rates of the birth-death process within the population depend on a discrete state $\sigma$ of the environment. We write $\omega_{i,\sigma}^\pm$ for the probability to transition from $i$ to $i\pm 1$ if the environment is in state $\sigma$. In principle, much of our theory applies to arbitrary number of environmental states, although our analysis of multi-player games focuses on the case of two such states.  Crucially, we assume that the process of the environmental variable is Markovian and that its transition rates do not depend on the state, $i$, of the population. We write $\mu_{\sigma\to\sigma'}$ for the probability that the state of the environment changes to $\sigma'$ in the next time step, if it is currently in state $\sigma$. Finally, we assume that the states of both the population and the environment can change in any one time step (i.e., the two processes are not mutually exclusive).

In our model the dynamics of the population occurs via a standard Moran process \cite{moran:book:1962,nowak:book:2006,traulsen:bookchapter:2009}
\BE
\omega_{i,\sigma}^+&=&\frac{i(N-i)}{N^2}\frac{f_A^\sigma(i)}{f^\sigma(i)}, \nonumber \\
\omega_{i,\sigma}^-&=&\frac{i(N-i)}{N^2}\frac{f_B^\sigma(i)}{f^\sigma(i)},
\EE
where  $f_A^\sigma(i)$ denotes the reproductive fitness of species $A$ in a population with $i$ individuals of type $A$ and when the environment is in state $\sigma$. The quantity
\be
f^\sigma(i)=\frac{i f_A^\sigma(i)+(N-i) f_B^\sigma(i)}{N}
\ee
is the mean fitness in the population if the environment is in state $\sigma$. The states $i=0$ and $i=N$ are absorbing states of the population dynamics for all environmental states, i.e. $\omega_{i=0,\sigma}^+=\omega_{i=N,\sigma}^-=0$. The above birth and death rates depend on the state of the environment, $\sigma$, through the fitnesses $f_A^\sigma(i)$ and $f_B^\sigma(i)$. These in turn are derived from payoffs in a multi-player game via the common exponential mapping  \cite{traulsen:bookchapter:2009}  $f_A^\sigma(i)=e^{\beta \pi_A^\sigma(i)},  f_B^\sigma(i)=e^{\beta \pi_B^\sigma(i)}$, where $\beta\geq 0$ denotes the intensity of selection, and where $\pi_A^\sigma(i)$ and $\pi_B^\sigma(i)$ are the payoffs to individuals of either type in the multi-player game in environment $\sigma$. 
\subsection{Multi-player games}
We assume that individual-based interactions take place between $n$ players, where $n$ is a fixed integer, $2\leq n\leq N$. We will write $a_{j,\sigma}$ for the payoff to an individual of type $A$ if they face $j$ other individuals of type $A$ and $n-1-j$ individuals of type $B$ in such an encounter of $n$ players. The payoff to a player of type $B$ is $b_{j,\sigma}$ is they play against $j$ individuals of type $A$, and $n-1-j$ (other) players of type $B$. These payoffs depend on the environmental state $\sigma$. 

This information is summarised in the following payoff matrix (cf. \cite{gamesmultiverse})
\be
\begin{tabular}{c|ccccc} 
 & $A\cdots A$ & $A\cdots AB$ & $\cdots$ & $AB\cdots B$& $B\cdots B$ \\
 \hline
$A$ & $a_{n-1,\sigma}$ & $a_{n-2,\sigma}$ &$\cdots$ & $a_{1,\sigma}$ & $a_{0,\sigma}$\\
$B$ & $b_{n-1,\sigma}$ & $b_{n-2,\sigma}$ &$\cdots$ & $b_{1,\sigma}$ & $b_{0,\sigma}$,\\
\end{tabular}
\ee
and we have $\pi_A^\sigma(i)=\sum_{j=0}^{n-1} a_{j,\sigma} H^A(i,j)$, and $\pi_B^\sigma(i)=\sum_{j=0}^{n-1} a_{j,\sigma} H^B(i,j)$. The quantity $H^A(i,j)$ is the probability that precisely $j$ individuals among the $n-1$ opponents of a player of type $A$ are of type $A$ as well, in a population with $i$ type $A$ individuals in total. Similarly, $H^B(i,j)$ is the probability for an individual of type $B$ to face precisely $j$ individuals of type $A$, and $n-1-j$ players of type $B$. Detailed expressions can for example be found in \cite{gamesmultiverse}.

In order to analyse the evolutionary dynamics of games in finite populations, it is often useful to first consider the limit of an infinite population. Stochastic effects are suppressed in this limit, but often the qualitative behaviour of the stochastic system is determined by the structure of the underlying deterministic flow. Writing $x = <i>/N$, where angular brackets denote an ensemble average, and keeping the environment fixed for the time being, one finds the following dynamics in the limit $N\to\infty$:
\be\label{eq:det}
\dot x = \omega_\sigma^+(x)-\omega_\sigma^-(x),
\ee
where $\omega_\sigma^\pm(x)$ is obtained from $\omega_{i,\sigma}^\pm$ by obvious substitutions. Eq. (\ref{eq:det}) can formally be derived from the leading order of an expansion in the inverse system size, or equivalently from the Kramers-Moyal expansion of the relevant master equation; see \cite{gardinerbook, kampenbook} for details. We will mostly be interested in the fixed points of Eq. (\ref{eq:det}). The rates $\omega_\sigma^{\pm}(x)$ contain a factor of $x(1-x)$. This is because evolutionary events affecting the state of the population require two individuals of the different types (one selected for reproduction, the other for removal). As a consequence, one always has the trivial fixed points at $x=0$ and $x=1$. Further fixed points can exist at locations $x^\star$ ($0\leq x^\star\leq 1$) for which $\pi_A^\sigma(x^\star)=\pi_B^\sigma(x^\star)$. The underpinning deterministic flows of multi-player games can then, qualitatively, be classified according to the number and stability of these internal fixed points.

Below we will focus on three-player games with a designated deterministic structure. We choose the values of the payoff matrix elements, $a_{j,\sigma}$ and $b_{j,\sigma}$, so as to have fixed points in specific locations; a prescription for doing this is given for general $n$-player games in the Supplement.  
\begin{figure}[t!!]
	\centering
	\includegraphics[scale = 0.5]{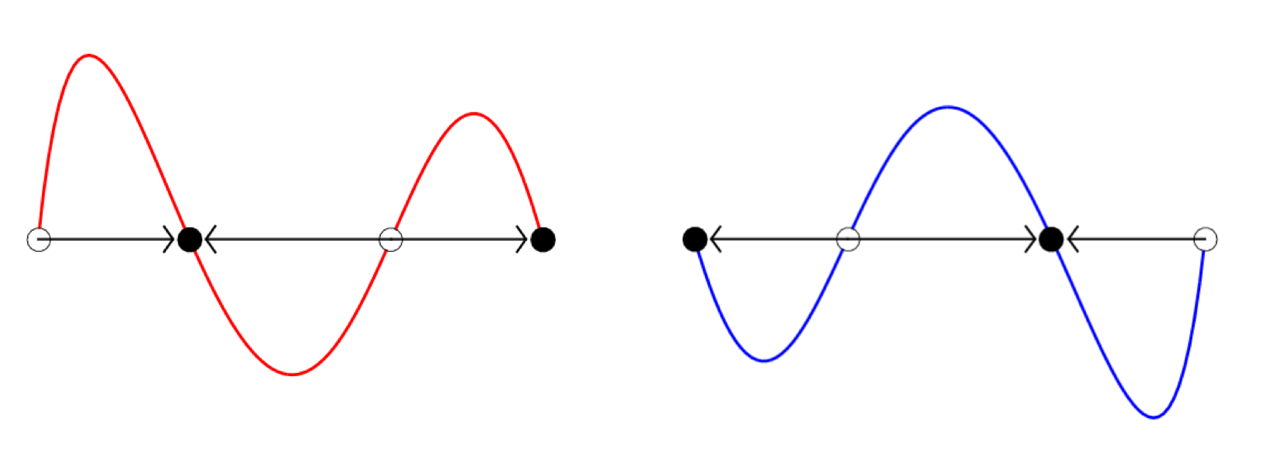}
	\caption{Deterministic flow in the two environments, we plot $\dot x$ as a function of $x$, see text for details. The internal fixed points are located at $x^\star=0.3$ and $x^\star=0.7$. The environment $\sigma = 1$ is shown on the left and $\sigma = -1$ is shown on the right. The fixed points have opposite stability in the two environments. The method of choosing payoff matrix elements such that fixed points are at designated locations is described in the Supplement.}
	\label{fig:xdotvsx}
\end{figure}

A typical example of the resulting deterministic flow can be found in Fig. \ref{fig:xdotvsx}, where plot $\dot x$ as a function of $x$. We show a case of a game with (non-trivial) fixed points at $x^\star=0.3$ and $x^\star=0.7$, but in which the deterministic flow has opposite directions in the two environments. In Fig. \ref{fig:xdotvsx} and in all subsequent figures below, we have always set $b_{j,\sigma} =1$ ($j=0,1,2$) in both environments $(\sigma=\pm 1)$. Effectively, this is an overall normalisation of payoff, individuals of the wildype $B$ always have $\pi_B\equiv 1$, irrespective of the composition of the population. The remaining pay-off matrix elements, $a_{j,\sigma}$ are then calculated as described in the Supplement.  We use $\beta=1/2$ in all sections except Sec. \ref{sec:finite}.

\section{Fixation probabilities, fixation times and sojourn times}\label{sec:sojourn}
\subsection{Fixation probability and fixation times}
We will now turn to the dynamics in finite populations, and consider situations in which one single mutant is placed in a population of $N-1$ wildtype individuals, i.e., the initial condition is $i=1$. While the deterministic flow of the dynamics in the limit of infinite populations may have internal fixed points, the outcome of evolution in a finite population will necessarily be extinction or fixation of the invading mutant; through genetic drift the population will eventually reach one of the absorbing states, $i=0$ (extinction of the mutant) or $i=N$ (fixation), and no further dynamics occurs. To analyse these processes we study the probability that the mutants fixates (as opposed to going extinct), and the mean time that they take to do so. 

The fixation probabilities and fixation times in switching environments have already been calculated in  \cite{ashcroft}, and we briefly summarise the results here.  Writing $\varphi_{i,\sigma}$ for the probability that the system reaches fixation ($i = N$), given that it currently contains $i$ individuals of type $A$ and is in environment $\sigma$, one has
\BE
\varphi_{i,\sigma}=\sum_{\sigma'}\mu_{\sigma\to\sigma'}\left[\omega_{i,\sigma'}^+ \varphi_{i+1,\sigma'}+\omega_{i,\sigma'}^- \varphi_{i-1,\sigma'}+(1-\omega_{i,\sigma'}^+-\omega_{i,\sigma'}^-)\varphi_{i,\sigma'}\right],\label{eq:phi}
\EE
subject to the boundary conditions $\varphi_{0,\sigma}=0$ and $\varphi_{N,\sigma}=1$ for all $\sigma$. Fixation times can be obtained from a similar relation. The unconditional fixation time $t_{i,\sigma}$ is the time until absorption (either at $i=0$ or at $i=N$) if the population starts in state $i$ and if the initial state of the environment is $\sigma$. Similarly, we can introduce the conditional fixation time for a given initial condition; this is an ensemble average of fixation time for trajectories that result in the fixation of the mutant. A procedure for solving Eqs. (\ref{eq:phi}) and its analogue for the conditional and unconditional fixation times in switching environments is described in detail in \cite{ashcroft}.  

\subsection{Sojourn times}
 We next introduce an additional tool for the analysis of population dynamics in switching environments, and describe the calculation of mean sojourn times. The concept of sojourn times is well known in the context of birth-death processes with absorbing states \cite{ewens:book:2004}; they describe the mean time spent in each state before absorption occurs. More precisely, we write $t_{i,\sigma;j}$ for the mean time the population spends in state $j$ before absorption, if it is started in state $i$ and if the initial state of the environment is $\sigma$. We stress that no requirement is made for the environment to be in a specific state when the population `sojourns' in $j$. Additionally, for the purposes of unconditional sojourn times, we do not specify whether the dynamics ends in the state with $0$ or $N$ mutants. The end-point is relevant for so-called conditional sojourn times though; we write $t_{i,\sigma;j}^\star$ for the mean time the population spends in state $j$, if started at $i$ and with initial environmental state $\sigma$, conditioned on fixation of the mutant. In simulations this object is measured as follows: Run a large number of independent realisations, all started from a population with $i$ mutants and $N-i$ wildtypes, and in environment $\sigma$. Then run each of these samples until the mutant has either gone extinct or reached fixation. Only take into account the runs in which fixation occurs. In this sample of trajectories measure the average time the population has spent in state $j$.
 
As part of this work we have developed a method for the calculation of conditional and unconditional sojourn times for general birth-death processes with two species and in switching environments. The calculation is broadly based on backward equation techniques \cite{gardinerbook, ashcroft}, and relies on `return probabilities', that is the likelihood that the population returns to state $i$ at a later time if it is started there. The key novelty here is that the dynamics of the enviroment needs to be accounted for as well. The calculation is lengthy, and we relegate the full mathematical details to the Supplement. The unconditional and conditional sojourn times discussed in the next section for specific games and environmental processes have been computed from Eqs. (S27) and (S34) respectively.

\section{Dynamics of three-player games}\label{sec:3playergames}

Using this analytical approach to calculate the sojourn times in switching environments (see Supplement), we can now examine the specific case of three-player games. We notice interesting behaviour in the three-player games which is not found in the two-player case; a minimum in the conditional fixation times with respect to the switching parameters for example. The sojourn times allow us to investigate where the system spends its time and to understand in more detail the mechanistic origin of the observed behaviour.
\begin{figure}[t!!]
	\centering
	\includegraphics[scale = 0.38]{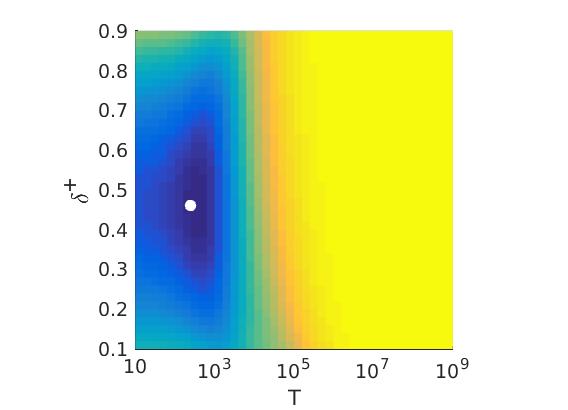}
	\includegraphics[scale = 0.38]{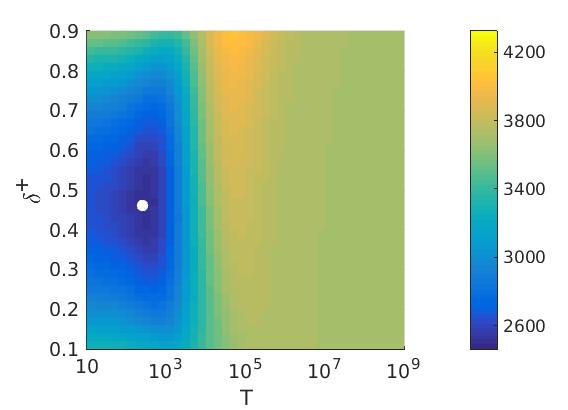}
	\caption{Conditional fixation time as a function of the switching parameters $T$ and $\delta^+$. Starting environment is $\sigma = 1$ in the left-hand panel and $\sigma = -1$ on the right. The fixed points of the system are at $x^\star = 0.3$ and $x^\star = 0.7$. Filled white dots in each panel indicate the point at which the conditional fixation time is minimal. }
	\label{fig:fixTime}
\end{figure}
\subsection{Notation}
We focus on systems in which the environment can take two different states, $\sigma=\pm 1$. In \cite{ashcroft} results for the case of two-player games were described as a function of the switching probabilities of the environmental state (per time step), $p^+=\mu_{+\to -}$ and $p^-=\mu_{-\to +}$. In order to better reflect the characteristic properties of the switching process we introduce $T = 1/p^+ + 1/p^-$ and $\delta^+ = p^-/(p^-+p^+)$, and use these quantities in place of the parameters $p^\pm$. The parameter $T$ can be interpreted as the average length of time that it takes for the environment to switch from one state to the other, and then back again. We will refer to $T$ as the `cycle time' or `switching period', keeping in mind though that the switching between the environments is stochastic so that there are no strictly periodic cycles. The parameter $\delta^+$ is the average proportion of time spent in the environment $\sigma = 1$. The proportion of time spent in environment $\sigma=-1$ is given by $\delta^-=p^+/(p^++p^-)=1-\delta^+$.  In order for $0\leq p^\pm \leq 1$, we require $T\geq 2$ and $1/T \leq \delta^+ \leq 1 - 1/T$. These conditions are easily understood, keeping in mind the discrete-time nature of the dynamics. The typical cycle must be at least two time steps long, and the minimum proportion of time spent in either environment is one time step, i.e. a fraction $1/T$ of the total (average) cycle.

We are most interested in the dynamics of the system when the internal fixed points are neither very close to one another, nor to the absorbing boundaries. Cases where the fixed points are close to one another or to the boundaries exhibit behaviour similar to games where there are only one or no internal fixed points. Novel behaviour is most clearly observed when the internal fixed points are sufficiently isolated, and so we focus on this regime.

In the following we will mostly use an example in which the two internal fixed points are located at $x^\star=0.3$ and $x^\star = 0.7$, and with deterministic flow as shown in Fig. \ref{fig:xdotvsx}. We have tested other cases and have observed similar behaviour.  Unless specified otherwise we always use a population size of $N=50$, and start with one single mutant $i=1$.

\subsection{Conditional fixation time}
The conditional fixation time is obtained using the formalism of \cite{ashcroft}, and is shown as a function of the switching parameters $T$ and $\delta^+$ in Fig. \ref{fig:fixTime}. The data in the figure reveals that there is a particular set of switching parameters $T$ and $\delta^+$ which minimises the conditional fixation time, as indicated by filled circles in Fig. \ref{fig:fixTime}. That is, we have a minimum with respect to $T$ for fixed $\delta^+$ and vice versa. In order to investigate the details of the dynamics leading to this effect, it is useful to discuss the conditional sojourn times, to gain an understanding of where the population spends its time on the way to fixation. These are obtained using the theory detailed in the Supplement, results are shown in Fig. \ref{fig:conditionalSojournTime}.

\begin{figure}[t!!]
	\centering
	\includegraphics[scale = 0.38]{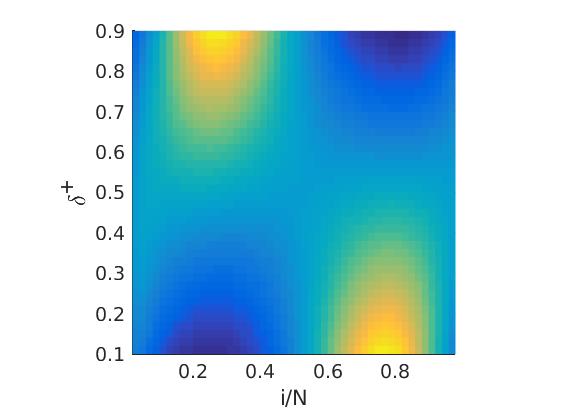}
	\includegraphics[scale = 0.38]{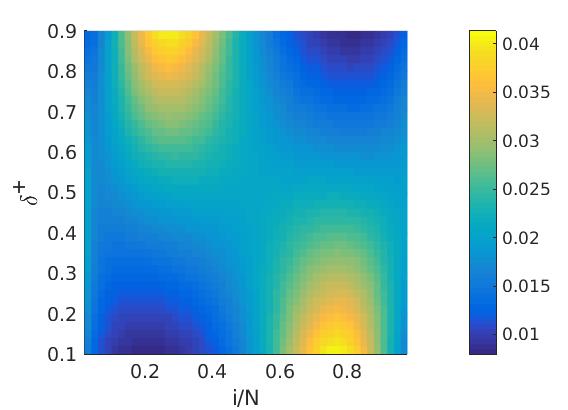}
	\caption{Re-scaled conditional sojourn time as a function of the switching parameter $\delta^+$ and the position $x = i/N$. Specifically we plot the fraction of time spent in each state, i.e., the sojourn times have been normalised to sum to unity for fixed $\delta^+$. Starting environment $\sigma = 1$ (left) and $\sigma = -1$ (right). We have fixed $T = 100$. When $\delta^+ \sim 1$ or $\delta^+ \sim 0$ the system tends to loiter around the stable fixed point in the $\sigma = 1$ and $\sigma = -1$ environments respectively. }
	\label{fig:conditionalSojournTime}
\end{figure}

We first focus on the effects of varying the model parameters $\delta^+$ and $\delta^-$, which reflect the proportion of time spent in each of the two environmental states. As seen in Fig. \ref{fig:conditionalSojournTime} successful trajectories (i.e. those in which the mutant reaches fixation) spend most of their time around the stable fixed point of the `dominant' environment, when either $\delta^+\gg \delta^-$, or vice versa. For example, when $\delta^+$ is close to one, most time is spent near the stable fixed point $x^\star = 0.3$ of environment $\sigma=1$. This illustrates that successful trajectories will loiter around that the relevant stable fixed point fixed point if one environmental state is significantly more frequent than the other. However, as the parameters $\delta^+$ and $\delta^-$ are moved away from the extremes, the time the system spends in the different states, $i=1,\dots,N=1$, is more evenly distributed, and the population has a lower propensity to get trapped near fixed points. At the value of $\delta^+$ which corresponds to the minimum in the conditional fixation time (see Fig. \ref{fig:fixTime}), the conditional sojourn times are fairly evenly spread across $i$ (Fig. \ref{fig:conditionalSojournTime}).  Trajectories then do not loiter around any of the internal fixed points, and successful runs reach fixation relatively quickly.
\\
\begin{figure}[t!!]
	\centering
	\includegraphics[scale = 0.38]{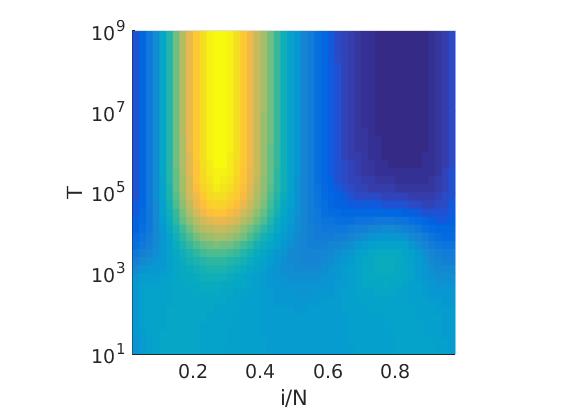}
	\includegraphics[scale = 0.38]{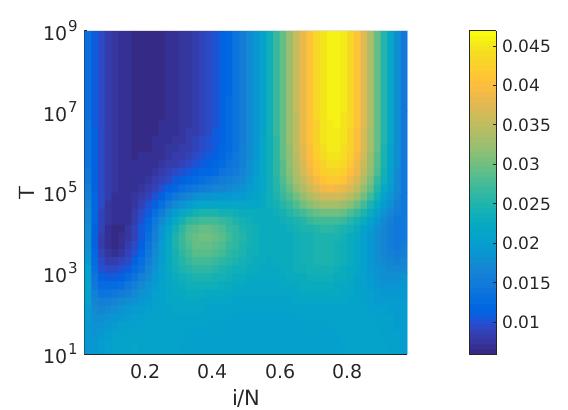}
	\caption{Re-scaled conditional sojourn time as a function of the switching parameter $T$ and the position $x = i/N$. Normalisation is as in the previous figure. Starting environment $\sigma = 1$ (left) and starting environment $\sigma = -1$ (right). We have fixed $\delta^+ = 0.5$.}
	\label{fig:conditionalSojournTimeVaryPeriod}
\end{figure}

We next turn to the effects of varying the time scale of the environmental switching time, i.e. the role of the parameter $T$. When the cycle period $T$ is small, the switching process between the environments is fast and so the initial state of the environment at the start of the dynamics does not have any significant effect. This is confirmed in Fig. \ref{fig:conditionalSojournTimeVaryPeriod}, where we show the conditional sojourn time as a function of $i/N$ and $T$ for fixed $\delta^+=\delta^-=1/2$. The two panels in the figure show the data for starts in the two different environments, and the sojourn times are seen to be quantitatively similar across the two panels when $T$ is small ($T\lesssim 10^3$ in our example). The sojourn time is then relatively constant across different states $i$ of the population, indicating that time is spent fairly evenly; we note again that the data in the figure is for $\delta^+ = 0.5$. This indicates that the population does not have sufficient time to settle near either of the fixed points, due to the fast switching dynamics. While each fixed point is attractive in one environment and a repeller in the other, these effects `average' out under fast environmental switching.
\par
For longer duration of the switching cycle $T$ however, the system begins to spend longer stretches of time in either environment, and so is able to spend more time at either fixed point. The conditional sojourn times begin to peak around the locations of the fixed points for $T \gtrsim 10^3$, see Fig. \ref{fig:conditionalSojournTimeVaryPeriod}.  If the cycle is longer still ($T\gtrsim 10^4$) most of the dynamics occurs in the starting environment, and so the population will spend most of its time around the stable fixed point in that environment. This is where we begin to see a marked difference between the two panels in Fig. \ref{fig:conditionalSojournTimeVaryPeriod}. For time scales above $T\approx 10^5$ the environment effectively never switches state before fixation is reached; this corresponds to the regime in Fig. \ref{fig:fixTime} in which the conditional fixation time reaches a value which is characteristic of the starting environment and independent of $\delta^+$.

The value of the cycle period $T\sim 10^3$ which minimises the conditional fixation time in Fig. \ref{fig:fixTime} corresponds roughly to the point at which the environment is neither switching so often that the trajectories are constantly changing direction, nor switching so little that the system loiters around a fixed point. This can be seen in Fig. \ref{fig:conditionalSojournTimeVaryPeriod} as the value of $T$ where the sojourn times begin to stop being spread equally across all values of $i/N$, and start to peak around both fixed points. 

\subsection{Probability for a single mutant to reach fixation}
\begin{figure}[t!!]
	\centering
	\includegraphics[scale = 0.37]{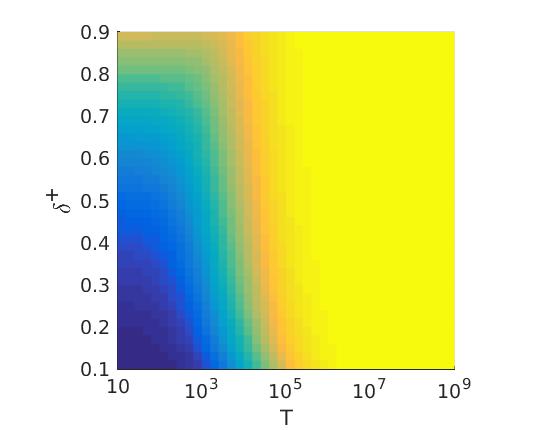}
	\includegraphics[scale = 0.38]{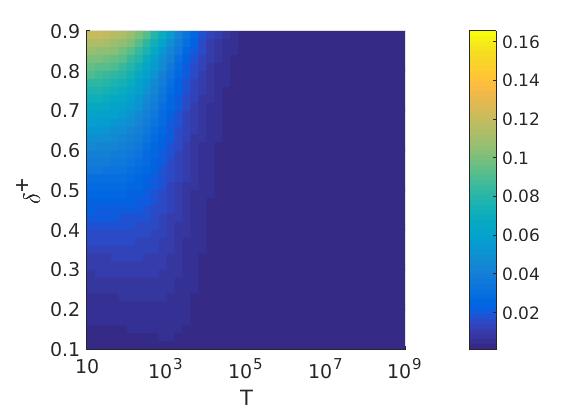}
	\caption{Fixation probability as a function of the switching parameters $T$ and $\delta^+$. Starting environment $\sigma = 1$ (left) and $\sigma = -1$ (right). Broadly speaking, the greater the amount of time the system spent in environment $\sigma = 1$, the greater the fixation probability.}
	\label{fig:fixProb}
\end{figure}

The probability that a single mutant reaches fixation is shown in Fig. \ref{fig:fixProb} as a function of the switching period, $T$, and the fraction of time spent in environment $\sigma=1$. The two panels show data for starts in either of the environmental states. The figure reveals that there is no combination of $T$ and $\delta^+$, which would extremise the fixation probability in the same way as the conditional fixation times. The main conclusion we draw from the figure, is, quite simply, that the likelihood that the mutant fixates is the greater the higher the proportion of time that the system spends in environment $\sigma=+1$, corresponding to the flow depicted in the left-hand panel of Fig. \ref{fig:xdotvsx}. This is the case for starts in either of the two environmental states. We also note that the fixation probability increases with the cycle period $T$ if the start occurs in environment $\sigma = 1$, but that it decreases with $T$ for starts in $\sigma = -1$.

\begin{figure}[t!!]
	\centering
	\includegraphics[scale = 0.37]{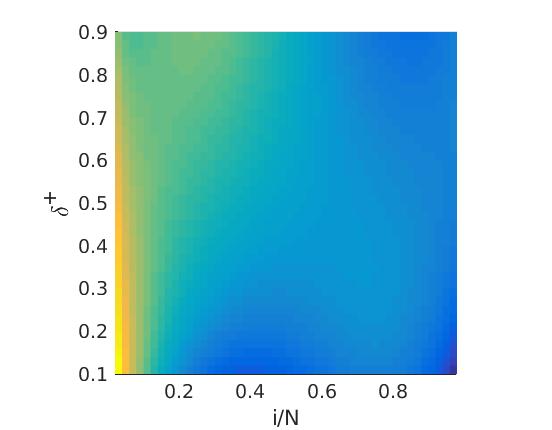}
	\includegraphics[scale = 0.37]{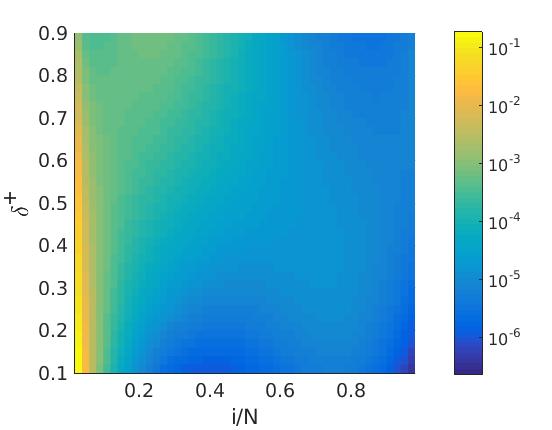}
	\caption{Unconditional sojourn time as a function of the switching parameter $\delta^+$ and the position $x = i/N$. Starting environment $\sigma = 1$ (left) and $\sigma = -1$ (right). Data is for fixed $T = 100$. The system can be seen to spend less time around the initial position at $i = 1$ as $\delta^+$ is increased, indicating that the system leaves the starting position more easily when $\delta^+$ is higher, as one might expect. This correlates with the increase in fixation probability with $\delta^+$ in Fig. \ref{fig:fixProb}. The qualitative behaviour is roughly the same in either environment.}
	\label{fig:unconditionalSojournTimes}
\end{figure}

\begin{figure}[t!!]
	\centering
	\includegraphics[scale = 0.37]{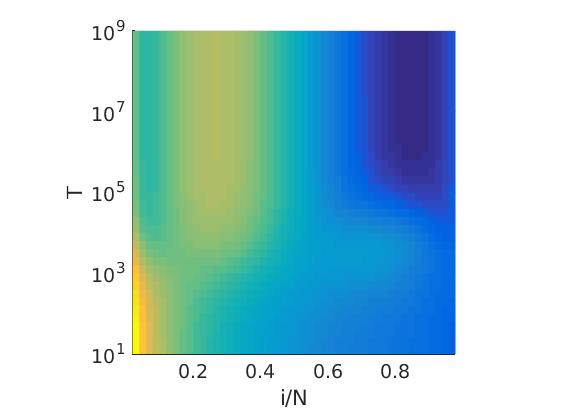}
	\includegraphics[scale = 0.37]{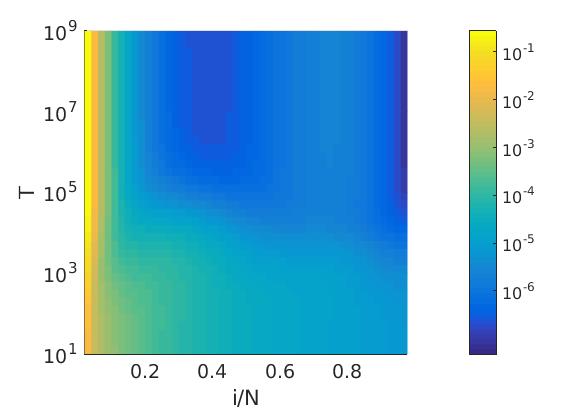}
	\caption{Unconditional sojourn times as a function of the switching parameter $T$ and the position $x = i/N$. Starting environment $\sigma = 1$ (left) and $\sigma = -1$ (right), for fixed $\delta^+ = 0.5$. As $T$ is increased, the system spends more time in the starting environment. In the left-hand panel, the system moves away from the starting position more easily with increasing $T$, which corresponds to the increase with $T$ of the fixation probability in the left-hand panel of Fig. \ref{fig:fixProb}. The opposite is true in the right-hand panel. The fixation probability can be seen to decrease with $T$ in the right-hand panel of Fig. \ref{fig:fixProb}.}
	\label{fig:unconditionalSojournTimesVaryPeriod}
\end{figure}

We now comment on the mechanisms generating these effects. The system always begins with one mutant, very close to the absorbing boundary at $i=0$. One might expect, therefore, that the propensity of the system to move away from the initial position, and from extinction, would have a great impact on the over all fixation probability. The unconditional sojourn times shown in Fig. \ref{fig:unconditionalSojournTimes} confirm that the system spends less of its trajectory near the starting position as $\delta^+$ is increased. This correlates with the increase in fixation probability with $\delta^+$ in Fig. \ref{fig:fixProb}. In Fig. \ref{fig:unconditionalSojournTimesVaryPeriod} we show the unconditional sojourn time as a function of the cycle period $T$. As this time scale $T$ is increased, the initial state of the environment becomes more relevant, as the first switch occurs later on average. This reduces the probability of immediate extinction on starts in the environment $\sigma = 1$ for which the flow is away from the $i=0$ boundary. This correlates with the increase in fixation probability with $T$ in the left-hand panel of Fig. \ref{fig:fixProb}. If the initial state of the environment is $\sigma=-1$ the argument is reversed. In this environment the gradient of selection is towards $i=0$ for small mutant numbers; a prolonged initial time spent in this environment hence reduces the likelihood that the mutant is successful, as indicated by the decrease of fixation probability in the right-hand panel of Fig. \ref{fig:fixProb}.

\subsection{Fixation in finite time}\label{sec:finite}
\begin{figure}[t!!]
	\centering
 \includegraphics[scale = 0.37]{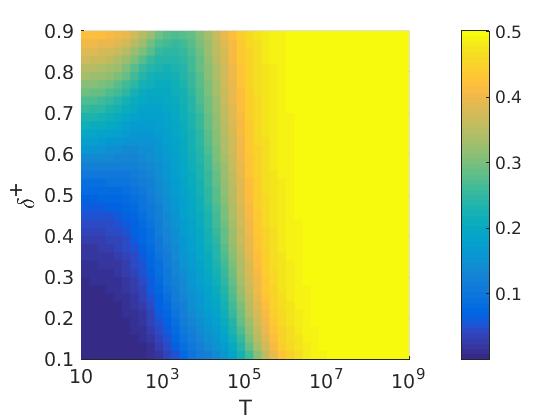}~~~
 \includegraphics[scale = 0.37]{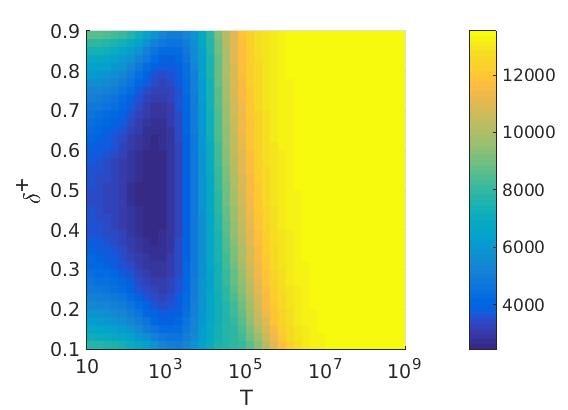}
 	\caption{Fixation probability (left) and conditional fixation time (right) as a function of the switching parameters $\delta^+$ and $T$ for the case $\beta=1$. In both cases, the starting environment is $\sigma = 1$ and the fixed points are located at $x^\star=0.3$ and $x^\star=0.7$. The behaviour is qualitatively the same as in Figs. \ref{fig:fixProb} and \ref{fig:fixTime}.}
	\label{fig:beta1}
\end{figure}

The combination of Figs. \ref{fig:fixTime} and \ref{fig:fixProb} reveals an intriguing observation. The fixation time, conditioned on fixation of the mutant, is minimal for a certain combination of $T$ and $\delta^+$, as indicated in Fig. \ref{fig:fixTime}. On the other hand, no such extremum is found for the fixation probability in Fig. \ref{fig:fixProb}. The fixation probability measures the likelihood for the mutant to be successful eventually -- including at long times. This indicates an interesting balance of two effects: if evolution is allowed to run indefinitely the mutant has the highest chance of success when $T$ and $\delta^+$ are large, and the starting environment is $\sigma = 1$. If however, we are interested in fixation on moderate time scales, the mutant does better near the point of minimal conditional fixation time in Fig. \ref{fig:fixTime}. In order to investigate this further we focus on a case with relatively strong selection, $\beta=1$. The corresponding conditional fixation times and fixation probabilities are shown as functions of $\delta^+$ and $T$ in Fig. \ref{fig:beta1}. Again a minumum in fixation time is found, but no extremum of the fixation probability. 

We introduce $Q_{i,\sigma}(t)$ as the probability that the mutant has reached fixation $t$ time steps (or sooner) after the system is started with $i$ mutants and in environmental state $\sigma$. We then have the discrete-time backward master equation (see e.g. \cite{gardinerbook,kampenbook,ewens:book:2004} for general references)
\be
Q_{i,\sigma}(t+1)=\sum_{\sigma'}\mu_{\sigma\to\sigma'}\left[\omega_{i,\sigma}^+ Q_{i+1,\sigma'}(t)+\omega_{i,\sigma}^- Q_{i-1,\sigma'}(t)+(1-\omega_{i,\sigma}^+-\omega_{i,\sigma}^-)Q_{i,\sigma'}(t)\right].\label{eq:bw}
\ee
The initial conditions are $Q_{i,\sigma}(t=0)=\delta_{i,N}$ for all $\sigma$, and allow us to numerically obtain the $Q_{i,\sigma}(t)$ by iterating Eq. (\ref{eq:bw}) forward.

\begin{figure}[t!!]
	\centering
	\includegraphics[scale = 0.37]{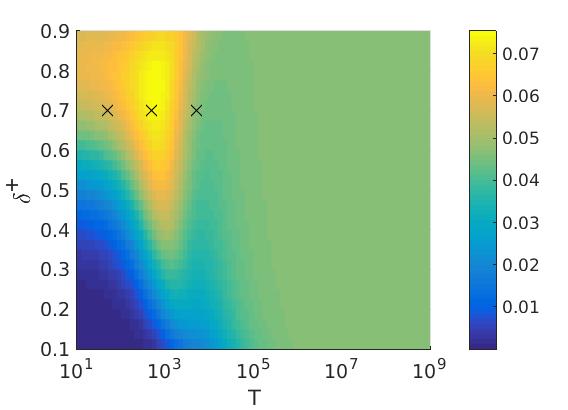}	~~~ \includegraphics[scale=0.32]{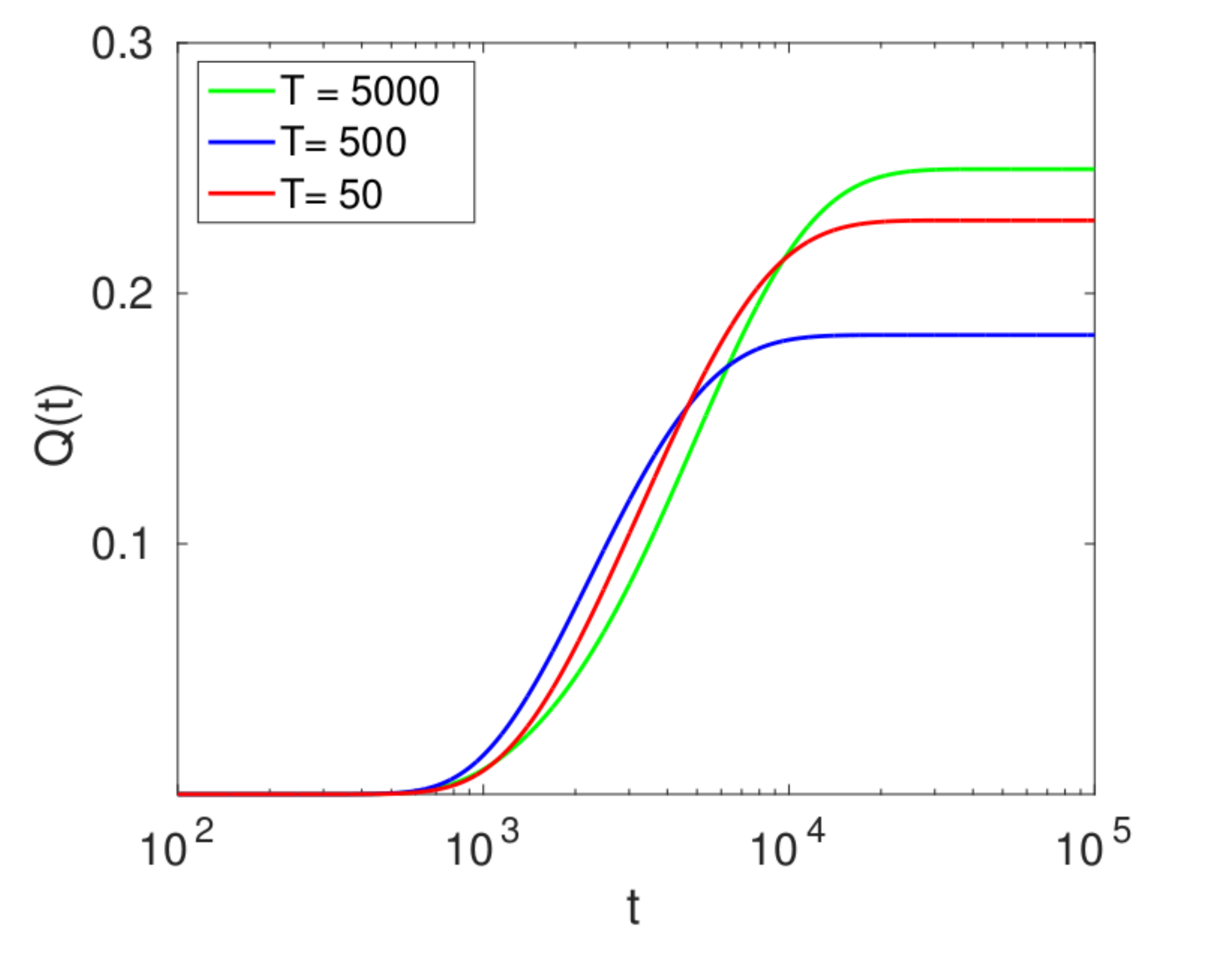}
 	\caption{Left: Proportion of trajectories which have reached state $i=N$ by time $t=2000$. Right: Fraction of trajectories which have reached $i=N$ by time $t$. We fix $\delta^+=0.7$ and show three different choices of $T$ in the right-hand panel; these are indicated by the crosses in the left-hand panel.  All data is for $\beta=1$, starting at $i=1$ and in environment $\sigma=1$.}
	\label{fig:finitet}
\end{figure}

Results are summarised in Fig. \ref{fig:finitet}. In the left-hand panel we show the fraction of samples, started with one single mutant, in which the mutant has reached fixation by time $t=2000$, a maximum is discernible near $\delta^+=0.7$ and just below $T\approx 10^3$. This demonstrates that the total fraction of samples that reach fixation by a finite time $t$ can have a local maximum at a location in the $\delta^+-T$ plane, even when the eventual fixation probability does not have such a maximum (cf. left-hand panel of Fig. \ref{fig:beta1}). This is clarified further in the right-hand panel of Fig. \ref{fig:finitet}, where we show $Q_{1,\sigma=1}(t)$ as a function of time, $t$, for three different choices of the cycle period (and at a fixed value of $\delta^+=0.7$). These correspond to the points marked in the left-hand panel of Fig. \ref{fig:finitet}. While more trajectories reach fixation eventually for $T=500$, the proportion that have reached fixation at times $t\approx 1000-5000$ is higher for the other two choices of $T$ shown in the figure ($T=50$ and $T=5000$). 
\section{Conclusions and outlook}\label{sec:outlook}
In summary, we have analysed the dynamics of fluctuating multi-player games in finite populations. Fluctuations of the payoff matrix are taken to originate from changes of an environmental state affecting the relative success of mutants and resident wildtypes. These environmental fluctuations could for example represent availability or absence of nutrients, the application of treatment in the context of bacterial populations or variations in other external conditions. Our analysis focuses on a stylised model with interaction between multiple individuals -- modelled as a multi-player game -- but representing general nonlinear payoff structures. This adds complexity relative to the two-player case studied in \cite{ashcroft}. In this earlier work fitness differences between mutants and wildtypes are linear in their relative frequencies, and so at most one internal equilibrium point is permitted by the resulting replicator equations. In this work we address higher-order interaction leading to more complex frequency-dependent fitness landscapes, with multiple non-trivial selection balance points. Specifically, we focus on games which allow two internal fixed points and on the regime in which these are well separated from each other and from the states at which mutants have reached fixation or gone extinct. We have then analysed in detail the likelihood for an invading mutant to reach fixation in the resident population. We have also studied the dynamics leading to fixation, in particular conditional fixation times, and the time spent in each state of the population on the path to fixation, the so-called sojourn times \cite{ewens:book:2004}. We have presented a comprehensive theory to calculate these analytically for multi-player games in finite populations subject to fluctuating environments with discrete states.

Our analysis  indicates that the time scales and detailed dynamics of the environmental switching process can affect the success of the invading mutant in several ways. For example, we find that the conditional fixation time can have a local minimum in the space of all (Markovian) switching processes, indicating that the mutant succeeds quickly under those circumstances, if it reaches fixation. At the same time the fixation probability appears to exhibit monotonous behaviour in the time-scale of the switching process and fraction of time spent in either of two environments. This indicates an interesting balance of two effects: a propensity to reach fixation quickly and  the overall fixation probability at long times. A more detailed analysis of the dynamics, based on a backward-master equation approach, reveals that these effects may be in competition with each other. There are switching dynamics for which there is a pronounced tendency to reach fixation in the early stages of the dynamics, but other parameters of the environmental dynamics may lead to a higher chance of fixation eventually.

In broader terms our findings contribute to constructing a more general theory of population dynamics with selection, random genetic drift and environmental fluctuations. While we have focused on a selected set of three-player games, the calculation of sojourn times in switching environments is applicable to general birth-death processes with fluctuating discrete environmental states. Hamilton has characterised the complexity of multi-player games as `sea-sickness' \cite{gamesmultiverse}. We believe our work is a contribution to reducing this discomfort in dealing with multi-player dynamics and towards an understanding of the outcomes of evolutionary processes with fluctuating non-linear interaction.

\subsection*{Acknowledgements}
We thank Pete(r) Ashcroft for discussions. JWB acknowledges a studentship by the Engineering and Physical Sciences Research Council (EPSRC, UK).

\renewcommand{\theequation}{S\arabic{equation}}
\renewcommand\thesection{S\arabic{section}}
\setcounter{equation}{0}
\setcounter{section}{0}
\newpage
{\bf \huge Supplementary Material}
\section{Construction of multi-player games}\label{app:gameconstruction}
We briefly describe the construction of multi-player games with specified locations of internal fixed points in the deterministic limit. In this limit the dynamics are given by
\be\label{eq:appdet}
\dot x = \omega_\sigma^+(x)-\omega_\sigma^-(x).
\ee
The subscript $\sigma$ is not relevant for the argument that follows, so we omit it for the remainder of this section of the Supplement. Given that each of the rates contains a factor $x(1-x)$ there are always two trivial fixed points $x^\star=0$ and $x^\star=1$. The remaining fixed points of Eq. (\ref{eq:appdet}) are the -- non-trivial -- roots of $\omega^+(x)-\omega^-(x)=0$. This is equivalent to 
\be
\pi_A(x)=\pi_B(x),
\ee
where the latter payoffs are given by 
\BE
\pi_A(x)&=&\sum_{j=0}^{n-1} a_j \left(\begin{array}{c} n-1 \\ j \end{array}\right) x^j (1-x)^{n-j-1} \nonumber \\
\pi_B(x)&=&\sum_{j=0}^{n-1} b_j \left(\begin{array}{c} n-1 \\ j \end{array}\right) x^j (1-x)^{n-j-1}
\EE
in an $n$-player game. We note that $\left(\begin{array}{c} n-1 \\ j \end{array}\right) x^j (1-x)^{n-j-1}$ is the probability for a given player to face $j$ opponents of type $A$ and $n-1-j$ opponents of type $B$ in an $n$-player game, if the fraction of type $A$-players in the population is $x$.
\subsection{General construction}
The construction of the game with specified non-trivial fixed points $x_1^\star,\dots, x_{n-1}^\star$ is equivalent to finding coefficients $c_j\equiv a_j-b_j$, $j=0,1,\dots,n-1$, such that
\be
\sum_{j=0}^{n-1} c_j \left(\begin{array}{c} n-1 \\ j \end{array}\right) x^j (1-x)^{n-j-1}=0
\ee
for $x\in\{x_1,\dots,x_{n-1}\}$.  

We construct a solution by induction. The problem is straightforward for $n=2$, in which case it reduces to finding $c_0$ and $c_1$ such that $c_0(1-x_1^\star)+c_1x_1^\star=0$. One finds
\be
\frac{c_{1}}{c_{0}} = 1 - \frac{1}{x_1^\star} .
\ee

Suppose now that we have constructed $c_0,\dots,c_{n-1}$ such that $x_1^\star,\dots,x_{n-1}^\star$ are the roots of
\be
\sum_{j=0}^{n-1} c_j \left(\begin{array}{c} n-1 \\ j \end{array}\right) x^j (1-x)^{n-j-1}=0.
\ee
We can introduce a new root at $x_n^\star$ by multiplying by $1- x + \left(1-\frac{1}{x_n^\star}\right)x$. One then obtains
\BE
&&\sum_{j=0}^{n-1}\left[c_{j}\left(\begin{array}{c} n-1\\ j\end{array}\right) + \left(1-\frac{1}{x_n^\star}\right) c_{j-1}\left(\begin{array}{c} n-1\\ j-1 \end{array}\right)\right]
x^j\left(1-x\right)^{n-j}\nonumber \\
&&+ c_{0}\left(1-x\right)^n
+ c_{n-1}\left(1-\frac{1}{x_n^\star}\right)x^n = 0 . 
\EE
This can be written in the form
\be
\sum_{j=0}^{n} c_j' \left(\begin{array}{c} n \\ j \end{array}\right) x^j (1-x)^{n-j}=0.
\ee
with
\BE
c_0' &=& c_0, \nonumber \\
c_n' &=& c_{n-1}\left(1-\frac{1}{x_n^\star}\right), \nonumber \\
c_{j}' &=& \frac{1}{\left(\begin{array}{c} n\\ j\end{array}\right)}
	\left[c_{j}\left(\begin{array}{c} n-1\\ j\end{array}\right)
	+c_{j-1}\left(1-\frac{1}{x_n^*}\right)\left(\begin{array}{c} n-1\\ j-1\end{array}\right)
	\right] . 
 \EE
This completes the inductive construction. 
\subsection{Three-player games}
The resulting relations are relatively compact for three-player games.  We will have internal fixed points at $x_1^\star$ and $x_2^\star$ if we choose
\BE
\frac{c_{1,\sigma} }{c_{0,\sigma} }= \frac{1}{2} \left[ \left(1-\frac{1}{x_1^\star} \right) + \left(1-\frac{1}{x_2^\star}\right) \right],
\label{eq:payoffs2}
\EE
and
\BE
\frac{c_{2,\sigma} }{c_{0,\sigma} } = \left( 1-\frac{1}{x_1^\star}\right)\left(1 - \frac{1}{x_2^\star}\right).
\label{eq:payoffs3}
\EE
We have here included the subscript $\sigma$ to indicate the dependence of the game on the environmental state. The coefficient $c_{0,\sigma}$ can be chosen arbitrarily, its sign determines the direction of the flow at $x=0$, and hence the stability of $x_1^\star$ and $x_2^\star$, respectively. In the main paper we use $c_{0, \sigma=1} = 1$ and $c_{0, \sigma=-1} = -1$.

 \section{Analytical calculation of sojourn times}
 \subsection{Initial steps of the calculation}
In order to compute sojourn times we introduce the quantity
\be
\varphi_{i,\sigma;j,\sigma'}=\mbox{Prob}\left(\begin{array}{p{7cm} | p{5cm}} There exists a time $t\geq t_0$ at which the system reaches state $j$, and when it does so for the first time the environment is in state $\sigma'$. & The population is started in state $i$ and the environent in $\sigma$ at initial time $t_0$. \end{array}\right).
\ee
We note that the initial time $t_0$ is obviously immaterial, as the dynamics is Markovian. One then has
\BE
\varphi_{i,\sigma;j,\sigma'}=\sum_{\sigma''} \mu_{\sigma\to\sigma''} \left[ \omega_{i,\sigma}^+ \varphi_{i+1,\sigma'';j,\sigma'}+\omega_{i,\sigma}^- \varphi_{i-1,\sigma'';j,\sigma'}+(1-\omega_{i,\sigma}^+-\omega_{i,\sigma}^-) \varphi_{i,\sigma'';j,\sigma'}\right],
 \label{eq:arrivalProbMaster}
 \EE
with the following boundary conditions
\be
\varphi_{0,\sigma;j,\sigma'}=\delta_{0,j}\delta_{\sigma,\sigma'}, ~
\varphi_{N,\sigma;j,\sigma'}=\delta_{N,j}\delta_{\sigma,\sigma'},  ~
\varphi_{j,\sigma;j,\sigma'}=\delta_{\sigma,\sigma'}.\label{eq:boundary}
\ee
The first and second of these reflect the fact that the states $i=0$ and $i=N$ are absorbing, so once the population is in state $i=0$ or $i=N$ it remains there at all future times. In the third relation in Eq. (\ref{eq:boundary}) we have $i=j$ so that trajectories, started at $(i,\sigma)$, will reach state $j$ immediately at $t=t_0$; they then only contribute to the above probability if $\sigma'=\sigma$.

For a fixed $j$ Eqs. (\ref{eq:boundary}) impose constraints on $\varphi_{i,\sigma;j,\sigma'}$ at $i=0$, $i=N$ and at $i=j$. Focusing on a given value of $j$ it is hence convenient to treat the cases $i<j$ and $i>j$ separately. To proceed we introduce the following quantities,
\be
\psi_{i,\sigma;j,\sigma'}=\sum_{\sigma''} \mu_{\sigma\to\sigma''}\varphi_{i,\sigma'';j,\sigma'}.
\ee
Focusing first on $j>i$, we define
\be
\nu_{i,\sigma';j,\sigma'}=\psi_{i,\sigma;j,\sigma'}-\psi_{i-1,\sigma;j,\sigma'},
\ee
and we then arrive at the following after substitution in Eq. (\ref{eq:arrivalProbMaster}),
\be
\nu_{i+1,\sigma;j,\sigma'}=\gamma_{i,\sigma}\nu_{i,\sigma;j,\sigma'}+\frac{1}{\omega_{i,\sigma}^+}\left[\left(\mumat^{-1}-\id\right)\sum_{k=1}^j \underline{\nu}_{k,\sigma;j}\right]_{\sigma'}. 
\ee
We have used vector and matrix notation for convenience. Indices run over the states of the environment; for example the components of $\underline{\nu}_{k,\sigma;j}$ correspond to the index $\sigma'$, and $\mumat$ has entries $\mu_{\sigma\to\sigma'}$. 

We then use the condition
\be
\sum_{i=1}^j\nu_{i,\sigma;j,\sigma'}=\mu_{\sigma\to\sigma'}
\ee
to determine $\nu_{i,\sigma;j,\sigma'}$. The probabilities $\varphi_{i,\sigma;j,\sigma'}$ can then be found using
\BE
\underline{\varphi}_{i,\sigma;j} = \mumat^{-1} \sum_{k=1}^i\underline{\nu}_{k,\sigma;j} .
\EE

Similarly, for $j<i$ we write $\lambda_{i,\sigma;j,\sigma} = \psi_{i,\sigma;j,\sigma} - \psi_{i+1,\sigma;j,\sigma}$, and find
\be
\lambda_{i-1,\sigma;j,\sigma'}=\frac{1}{\gamma_{i,\sigma}}\lambda_{i,\sigma;j,\sigma'}+\frac{1}{\omega_{i,\sigma}^-}\left[\left(\mumat^{-1}-\id\right)\displaystyle\sum_{k=i}^{N-1} \underline{\lambda}_{k,\sigma;j}\right]_{\sigma'}, 
\ee
One then proceeds using $\displaystyle\sum_{i=j}^{N-1}\lambda_{i,\sigma;j,\sigma'} = \mu_{\sigma\to\sigma'}$, and $\underline{\varphi}_{i,\sigma;j} = \mumat^{-1} \displaystyle\sum_{k=i}^{N-1}\underline{\lambda}_{k,\sigma;j}$.

Now we have at our disposal a means by which to calculate the complete set of probabilities $\varphi_{i,\sigma;j,\sigma'}$. Using these probabilities, one can then compute the unconditional and conditional sojourn times. 

\subsection{Calculation of unconditional sojourn times}
As the next step we compute

\be
r_{i,\sigma;\sigma'}=\mbox{Prob}\left(\begin{array}{p{6.5cm}|p{5cm}} There exists a time $t>t_0$ at which the population returns to state $i$, and when it does so for the first time the environment is in state $\sigma'$. & The population is started in state $i$ and the environent in $\sigma$ at initial time $t_0$. \end{array}\right).
\ee
 We stress the requirement that $t$ be strictly greater than $t_0$, marking a difference compare to the above definition of $\varphi_{i,\sigma;j,\sigma'}$, where we only require $t\geq t_0$. Hence, $r_{i,\sigma;\sigma'}$ is in general distinct from $\varphi_{i,\sigma;i,\sigma'} = \delta_{\sigma, \sigma'}$. We have
\BE
r_{i,\sigma;\sigma'} = \mu_{\sigma\to\sigma'}\left(1-\omega_{i,\sigma}^+-\omega_{i,\sigma}^-\right)+\sum_{\sigma''}\mu_{\sigma\to\sigma''}\left(\omega_{i,\sigma}^+\varphi_{i+1,\sigma'';i,\sigma'}+\omega_{i,\sigma}^-\varphi_{i-1,\sigma'';i,\sigma'}\right).
\label{eq:returnProb}
\EE
We can now turn to sojourn times. Consider a trajectory which begins in state $\left(i,\sigma\right)$. The probability of spending a total of $t$ time steps in a particular state $j$, irrespective of the state the environment is in at that time, is then given by 
\BE
q_t(j|i,\sigma) = \sum_{\sigma_{1}...\sigma_{t-1}} \varphi_{i,\sigma;j,\sigma_{1}} r_{j,\sigma_{1};\sigma_{2}}r_{j,\sigma_{2};\sigma_{3}}...r_{j,\sigma_{t-2};\sigma_{t-1}}\left(1-\sum_{\sigma_t}r_{j,\sigma_{t-1};\sigma_{t}}\right) .
\EE
The trajectory first has to reach state $j$, as indicated by $\varphi_{i,\sigma;j,\sigma_{1}}$, it then has to `return' $t$ times [in the sense of Eq. (\ref{eq:returnProb})], indicated by the factors $r_{j,\sigma_{1};\sigma_{2}}r_{j,\sigma_{2};\sigma_{3}}...r_{j,\sigma_{t-2};\sigma_{t-1}}$, and it must then not return to $j$ again, see the factor $1-\sum_{\sigma_t}r_{j,\sigma_{t-1};\sigma_{t}}$. This can be written in a more compact matrix notation
\BE
\underline{q_t}(j|i) = \underline{\underline{\varphi_{ij}}}\left(\underline{\underline{r_{j}}}\right)^{t-1}\underline{x_j},
\EE
where
\BE
\left(\underline{x_{j}}\right)_{\sigma} = 1 - \sum_{\sigma'}r_{j,\sigma;\sigma'} .
\EE
The unconditional sojourn time is then the first moment of the distribution over $t$ defined by $q_t(j|i,\sigma)$,
\BE
t_{i,\sigma;j} = \sum_{t=1}^\infty tq_{t}\left(j|i\sigma\right)=\left(\underline{\underline{\varphi_{ij}}}\left[\sum_{t=1}^\infty t\left(\underline{\underline{r_{j}}}\right)^{t-1}\right]\underline{x_j}\right)_{\sigma}.
\EE
Letting $\underline{\underline{\id}} - \underline{\underline{r_{j}}} = \underline{\underline{S_{j}}} $, one can evaluate the series to find
\BE
t_{i,\sigma;j} = \left(\underline{\underline{\varphi_{ij}}}\left(\underline{\underline{S_{j}^{-1}}}\right)^2\underline{x_j}\right)_{\sigma}.
\EE
Therefore, one can calculate the unconditional Sojourn times once the probabilities $\underline{\underline{\varphi_{ij}}}$ have been obtained as described above.

\subsection{Conditional sojourn times}
The conditional Sojourn times can be calculated in a similar way. We introduce the following shorthand
\be
q_{t}^*\left(j|i,\sigma\right) = \mbox{Prob}\left(\begin{array}{p{5cm}|p{5cm}} The population spends exactly  $t$ steps at $j$ before absorption. & The starting point is $(i, \sigma)$, and the mutant reaches fixation.\end{array}\right)  .
\ee

Using Bayes' theorem we have
\BE
q_{t}^*\left(j|i,\sigma\right) = \frac{\mbox{Prob}\left(\mbox{spends $t$ steps at $j$ and reaches fixation}~ |~ \mbox{starts at $(i, \sigma)$}\right)}
{\mbox{Prob}\left(\mbox{reaches fixation}~ |~ \mbox{starts at $(i, \sigma)$}\right)} .
\EE
We also note that 
\be
\mbox{Prob}\left(\mbox{reaches fixation}~ |~ \mbox{starts at $(i,\sigma)$}\right) = \varphi_{i,\sigma} = \sum_{\sigma'} \varphi_{i,\sigma ; N , \sigma'} .
\ee
Hence, similar to the unconditional case,
\be
q_{t}^*\left(j|i,\sigma\right) \sum_{\sigma'} \varphi_{i,\sigma ; N , \sigma'}
= \sum_{\sigma_{1}...\sigma_{t-1}} \varphi_{i,\sigma;j,\sigma_{1}} r_{j,\sigma_{1};\sigma_{2}}r_{j,\sigma_{2};\sigma_{3}}...r_{j,\sigma_{t-2};\sigma_{t-1}}\left(\underline{y_{j}}\right)_{\sigma_{t-1}},
\ee
where
\be
\left(\underline{y_{j}}\right)_{\sigma} = \omega_{j, \sigma}^+ \sum_{\sigma_{t}} 
\mu_{\sigma\to\sigma_{t}}\left(1 - \sum_{\sigma'} \varphi_{j+1,\sigma_{t} ; j, \sigma'}\right).
\ee

Using more compact notation we have
\BE
q_{t}^*\left(j|i,\sigma\right) = \frac{\left(\underline{\underline{\varphi_{ij}}}\left(\underline{\underline{r_{j}}}\right)^{t-1}\underline{y_j},\right)}{\displaystyle\sum_{\sigma'} \varphi_{i,\sigma ; N , \sigma'}},
\EE
Using a procedure similar to the unconditional case, we obtain
\BE
t_{i,\sigma;j}^* = \sum_{t=1}^\infty tq_{t}^*\left(j|i,\sigma\right) = \frac{\left(\underline{\underline{\varphi_{ij}}}\left(\underline{\underline{S_{j}^{-1}}}\right)^2\underline{y_j}\right)_{\sigma}}
{\displaystyle\sum_{\sigma'} \varphi_{i,\sigma ; N , \sigma'}} .
\EE

These conditional Sojourn times can also be calculated, given the complete set of probabilities $\varphi_{i,\sigma;j,\sigma'}$. 
\end{document}